\documentclass[]{aastex631}

\usepackage{amsmath}
\usepackage{mathtools}
\usepackage{booktabs}

\shortauthors{Hatta and Sekii}
\graphicspath{{./}{figures/}}

\begin{document}

\title{High-order Gravity-mode Period Spacing Patterns of \\
Intermediate-mass ($1.5 \, M_\odot < M < 3 \, M_{\odot}$) Main-sequence Stars I. Perturbative Analysis
}

\correspondingauthor{Yoshiki Hatta}
\email{yoshiki.hatta@isee.nagoya-u.ac.jp}


\author[0000-0003-0747-8835]{Yoshiki Hatta}
\affiliation{Institute for Space-Earth Environmental Research, Nagoya University \\
Furo-cho, Chikusa-ku, Nagoya, Aichi 464-8601, Japan}
\affiliation{Max-Planck Institute for Solar System Research \\
Justus-von-Liebig-Weg 3, 37077 G\"{o}ttingen, Germany}
\affiliation{National Astronomical Observatory of Japan \\
2-21-1 Osawa, Mitaka, Tokyo 181-8588, Japan}

\author[0000-0001-6583-2594]{Takashi Sekii}
\affiliation{National Astronomical Observatory of Japan \\
2-21-1 Osawa, Mitaka, Tokyo 181-8588, Japan}
\affiliation{Astronomical Science Program, The Graduate University for Advanced Studies, SOKENDAI\\
2-21-1 Osawa, Mitaka, Tokyo 181-8588, Japan}



\begin{abstract}
Theoretical study of high-order gravity-mode period spacing ($\Delta P_g$) pattern is relevant for the better understanding of internal properties 
of intermediate-mass ($1.5 \, M_\odot < M < 8 \, M_{\odot}$) main-sequence g-mode pulsators. 
In this paper, we carry out the first-order perturbative analysis to evaluate effects of a sharp, though not discontinuous, transition in the Brunt-V\"{a}is\"{a}l\"{a} (BV) frequency on the $\Delta P_g$ pattern. 
Such a finite-width transition in the BV frequency, whose scale height can be comparable to the local wavelength of gravity waves, is expected to develop in relatively low-mass ($1.5 \, M_\odot < M < 3 \, M_{\odot}$) main-sequence stars, causing a bump in the second derivative of the BV frequency. 
Inspired by Unno et al.'s formulation, we treat 
the bump in the second derivative of the BV frequency as a small perturbation, which allows us to derive an analytical expression of the $\Delta P_g$ pattern. 
The analytical expression shows that the amplitude of the oscillatory $\Delta P_g$ pattern is determined by a weighted average of the bump in the second derivative of the BV frequency where the weighting function 
is given by the g-mode eigenfunction. 
Tests with low-mass ($\sim 2 \, M_\odot$) main-sequence stellar models show that the analytical expression can reproduce the $\Delta P_g$ patterns numerically computed reasonably well. 
The results of our perturbative analysis will be useful for, e.g., improving semi-analytical expressions of the $\Delta P_g$ pattern, which would enable us to investigate $\Delta P_g$ patterns of SPB stars and $\gamma$ Dor stars for inferring chemical composition profile and rotation rates. 
\end{abstract}

\keywords{Asteroseismology (73) --- Stellar oscillations (1617) --- Stellar interiors (1606) --- Stellar cores (1592) --- Stellar evolution (1599)}


\section{Introduction} \label{sec:intro}
Modern space-borne missions such as Kepler \citep{2010ApJ...713L..79K} and TESS \citep{2014SPIE.9143E..20R} have enabled us to observe tens of thousands of stars that are oscillating with gravity (g) modes whose restoring force is the buoyancy. 
It is generally considered that g modes are formed in the deep radiative region of the stars \citep[e.g.][]{1989nos..book.....U}, and thus, g-mode detections are fairly useful for investigating internal dynamics and structure of the g-mode pulsators, 
leading to valuable constraints on stellar internal physics \citep[see][for the thorough review of the current status of g-mode asteroseismology]{2024A&A...692R...1A}. 

One observational feature frequently found in the g-mode pulsators' spectra is the so-called g-mode period spacing (hereafter $\Delta P_{g}$) pattern; when we take subtraction between two g-mode periods with the neighboring radial order ($n$) with the same spherical degree ($\ell$) and plot the period spacings as a function of the g-mode period, we see some oscillatory patterns around a general trend \citep[e.g.][]{2010Natur.464..259D,2014A&A...570A...8P,2015ApJS..218...27V,2020MNRAS.491.3586L,2022A&A...662A..82G,2025A&A...695A.214A}. 
Behaviors of the $\Delta P_g$ pattern can be explained based on asymptotic theory of stellar nonradial oscillation that focuses on modes with high radial orders \citep[roughly speaking, $n > 10-20$;][]{1979PASJ...31...87S,1980ApJS...43..469T}. 
In particular, the oscillatory patterns are thought to originate from mode trapping, in which some g modes are selectively trapped in a certain region of the star due to sharp variations in the Brunt-V\"{a}is\"{a}l\"{a} frequency (hereafter BV frequency) \citep[e.g.,][]{1992ApJS...80..369B,2003MNRAS.344..657M,2008MNRAS.386.1487M}. 
As such, the observed $\Delta P_g$ patterns enable us to infer the BV frequency profile in the deep radiative region of the stars. 
Importantly, what contributes to the BV frequency in the deep radiative region of stars is the chemical composition gradients that are determined by various mixing mechanisms inside \citep[e.g.,][]{2008MNRAS.386.1487M}. 
The oscillatory $\Delta P_g$ pattern is thus key to inferring the stellar internal mixing processes \citep[e.g.,][]{2022ApJ...930...94P}. 

One of the most common ways to estimate the internal chemical composition profile based on the observed $\Delta P_g$ pattern is the stellar modeling approach where a stellar evolutionary model that can reproduce the observed $\Delta P_g$ pattern as well as possbile is constructed. 
Nevertheless, especially in the case of intermediate-mass main-sequence g-mode pulsators such as SPB stars \citep[][]{1991A&A...246..453W} and $\gamma$ Dor stars \citep[e.g.,][]{2000ApJ...542L..57G} with the mass range of $3-8 \, M_\odot$ and $1.3-1.7 \, M_\odot$, respectively, the number of the attempts of the stellar modeling approach 
is small ($< 50$) \citep[e.g.,][]{2015A&A...580A..27M,2016ApJ...823..130M,2018A&A...616A.148B,2020ApJ...899...38W,2021A&A...650A.175M,2023A&A...679A...6M,2018A&A...614A.128P,2021NatAs...5..715P,2022ApJ...930...94P} compared with the large number of observed stars in this mass range that exhibit the $\Delta P_g$ pattern \citep[$> 1000$;][]{2025A&A...695A.214A}. 
This is mainly due to the fast rotation typical for early-type main-sequence stars 
\citep[the rotational period of about 1 day;][]{1995ApJS...99..135A,2009LNP...765..207R}; we have some difficulties in constructing a stellar model where rotational effects on the internal structure is taken into account self-consistently \citep[e.g.,][]{2022FrASS...9.4579R}. 
Besides, irrespective of whether the stellar modeling is carried out in 1-d or 2-d, 
we rely on mixing prescriptions in stellar evolutionary codes that are described phenomenologically with some free parameters; we need to somehow calibrate the mixing prescriptions for each of target stars. 
This difficulty has for instance led to an alternative approach where calibration is carried out with an ensemble of $\gamma$ Dor stars \citep{2024A&A...685A..21M}.  

In this context, the semi-analytical approach \citep{2015ApJ...805..127C,2019MNRAS.490..909C,2024A&A...687A.100C,2023ApJ...950..165H} may allow us to analyze the $\Delta P_g$ patterns of the SPB and $\gamma$ Dor stars in rather a model-independent manner \citep[e.g.,][]{2018A&A...618A..47C}. 
Pioneering work has been carried out by \citet{2019MNRAS.490..909C} (hereafter C19), where they applied the JWKB analysis to explicitly derive a semi-analytical expression of the $\Delta P_g$ pattern based on the two-zone modeling of the BV frequency profile. 
While C19 focus on the $\Delta P_g$ pattern of main-sequence ($\sim 6 \, M_\odot$) and red-giant ($\sim 1 \, M_\odot$) stars,  \citet{2023ApJ...950..165H} (hereafter H23) has expanded upon C19 to derive the semi-analytical expression for the $\Delta P_g$ pattern where the BV frequency has been modeled by a ramp function that (qualitatively) resembles the BV frequency of intermediate-mass main-sequence stellar models. 
H23's attempts are successful in reproducing the decreasing trends (as a function of the g-mode period) in the amplitude of the $\Delta P_g $ pattern of stellar models with the masses $> 3 \, M_{\odot}$, highlighting the potential that the semi-analytical expression can be applied to the SPB and $\beta$ Cep stars. 

Still, for lower-mass models ($M < 3 \, M_{\odot}$), H23's expression underestimates the period dependence of the amplitudes of the $\Delta P_g$ patterns. 
Since the majority of the intermediate-mass main-sequence g-mode pulsators are $\gamma$ Dor stars \citep[e.g.,][]{2020MNRAS.491.3586L} with the mass of about $1.3-1.7 \, M_\odot$, whether or not the semi-analytical approach can be refined to be applicable to the lower-mass models ($M < 2 \, M_\odot$) is crucial, which motivates us to improve H23's expression. 
Related to the motivation, it should be worth noting that inference on the internal rotation rates of the SPB and $\gamma$ Dor stars based on the observed $\Delta P_g$ patterns has been quite successful \citep[e.g.,][]{2025A&A...695A.214A}. 
Although the fast rotation of the SPB and $\gamma$ Dor stars renders them to be difficult targets to model \citep{2022FrASS...9.4579R}, the so-called traditional approximation of rotation \citep[TAR;][]{eckart1960hydrodynamics,1997ApJ...491..839L} allows us to evaluate the rotational effects on the $\Delta P_g$ pattern \citep{2013MNRAS.429.2500B}, resulting in more than hundreds of inferences on rotation rates around the deep radiative region of SPB and $\gamma$ Dor stars \citep[e.g.,][]{2019MNRAS.487..782L,2020MNRAS.491.3586L,2025A&A...695A.214A}. 
The validity of TAR has been mostly confirmed by the numerical computations of 2-d stellar oscillations \citep{2017MNRAS.465.2294O}, thus leading to the firm consensus that the deep radiative region of the intermediate-mass main-sequence stars is rotating as fast as the surface \citep{2024A&A...692R...1A}. 
We would expect similar progresses for inference on the chemical composition profile of the SPB and $\gamma$ Dor stars by refining the semi-analytical approach to describe the $\Delta P_g$ pattern. 

One probable cause of the discrepancy between H23's semi-analytical expression and the $\Delta P_g$ patterns of lower-mass ($M < 2 \, M_\odot$) main-sequence models is the ``finite-wavelength effect''; that is, how a g mode senses a sharp BV frequency variation with a finite width depends on the (local) wavelength of the g mode 
\citep[see, e.g., C19,][and H23 for more discussions]{2024A&A...687A.100C}. 
In H23, such a finite-wavelength effect is neglected, and (the first-derivative of) the BV frequency is assumed to transition discontinuously, which is suspected to be unphysical in the case of the stars. 
There is thus room for improving H23's expression by taking the finite-wavelength effect into account. 
In the case of red-giant stars, such attempts have already been taken by C19 based on that the finite-width BV frequency transition in red-giant stars can be modeled as a Gaussian-like glitch (see, e.g., panel d of Figure 1 in C19). 
Their semi-analytical expression can reproduce the $\Delta P_g$ patterns of red-giant stars very well, though they also see some systematics in recovering the glitch information \citep[see C19 and][for more discussions]{2024A&A...687A.100C}. 
It should also be noted that the BV frequency transition in the intermediate-mass main-sequence stars is not shaped like a Gaussian glitch as in the case of red-giant stars, requiring an approach different from C19's to evaluate the finite-wavelength effect seen in the intermediate-mass main-sequence stars. 

As a first step for the improvement on the semi-analytical expression for the $\Delta P_g$ pattern of main-sequence g-mode pulsators, we in this paper carry out the perturbative analysis of $\Delta P_g$ pattern with the purpose of understanding and interpreting the finite-wavelength effect. 
There have already been a number of literatures where the perturbative analysis of the $\Delta P_g$ pattern has been conducted \citep[e.g.,][]{2003MNRAS.344..657M,2008MNRAS.386.1487M,2020ApJ...899...38W}. 
But almost all of the previous studies treated the variation in the BV frequency as a small perturbation, and thus, we cannot apply the perturbation theory to explain $\Delta P_g$ patterns of stellar models in which the BV frequency can vary from the peak to ground values by factor a few. 
Very recently, \citet{2025arXiv251105780G} has carried out the perturbative analysis where the first derivative of the BV frequency is treated as a small perturbation, showing some success in extracting information on the BV frequency transition. 
In this paper, we will explain that $\Delta P_g$ patterns of stellar models with the mass of $\sim 2 M_{\odot}$ can be described by the perturbative approach if we treat the $f_1$-term, which corresponds to a bump in the second derivative of the BV frequency (later defined in Section \ref{sec:2}), as a small perturbation. 
The results of this paper are to be used for deriving a new semi-analytical expression for the $\Delta P_g$ pattern (the second of this series of papers; hereafter paper 2). 

This paper is organized as below. 
In Section \ref{sec:2}, we describe the first-order perturbative analysis to derive an analytical expression of the $\Delta P_g$ pattern. 
The analytical expression is subsequently validated using toy models of the BV frequency (Section \ref{sec:3}). 
Keeping in mind a possible application to structure inversion, we also test the analytical expression derived in Section \ref{sec:2} using the stellar models (Section \ref{sec:4}). 
We will conclude and give some future prospects in Section \ref{sec:5}. 

\section{Perturbative analysis} \label{sec:2}
In this section, based on the perturbation theory, we derive an analytical expression that relates the $\Delta P_g$ pattern to a sharp, though finite-width, transition in the BV frequency profile. 
We start with an equation that we would like to solve (Section \ref{sec:2-1}), and we describe the ``$f$-term'' that is introduced by \citet{1989nos..book.....U} (Section \ref{sec:2-2}). 
To carry out the perturbative analysis, we then decompose the equation into an unperturbed part and a perturbed part (Section \ref{sec:2-3}). 
After we show that the variational principle holds in this case (Section \ref{sec:2-4}), we relate the small perturbation (the ``$f_1$-term'') to the frequency perturbation (Section \ref{sec:2-5}), based on which we can compute the $\Delta P_g$ patterns. 
The analytical expression derived based on the perturbation analysis will be later validated in Sections \ref{sec:3} and \ref{sec:4}. 


\subsection{Equation} \label{sec:2-1}
The starting point of our perturbative analysis is the second-order differential equation, which 
is derived based on the Cowling approximation \citep[see Equation (16.11) in][hereafter U89]{1989nos..book.....U}: 
\begin{eqnarray}
	\frac{\mathrm{d^2}v}{\mathrm{d}r^2} 
	+ 
	[k_r^2 - \epsilon] v = 0, \label{eq:rev1}
\end{eqnarray}
where $r$ and $v$ are the radial coordinate and the radial displacement that is scaled by the structural variables (see Equation (16.9) of U89), respectively. 
The radial wavenumber $k_r$ is given by: 
\begin{eqnarray}
	k_r^2 = \frac{1}{\omega^2 c^2} (\omega^2 - N^2) (\omega^2 - L_\ell^2), \label{eq:rev2}
\end{eqnarray} 
in which $\omega$, $N$, and $L_\ell$ are the eigenfrequency, the BV frequency, and the Lamb frequency, respectively. 
The squared sound speed is denoted by $c^2$ with which the Lamb frequency is expressed as $L_{\ell}^2 = L^2 c^2 / r^2$. 
The constant $L$ is related to the degree of the spherical harmonics $\ell$ as $L = \sqrt{\ell (\ell+1)}$. 
The term $\epsilon$ is related to the second derivative of the structural variables, which is defined as below: 
\begin{eqnarray}
	\epsilon 
	\equiv 
	|P|^{1/2} 
	\frac{\mathrm{d}^2 |P|^{-1/2}}{\mathrm{d} r^2}, \label{eq:rev3}
\end{eqnarray}
where $P$ is expressed as (Equation (16.3) in U89): 
\begin{eqnarray}
	P = \frac{r^2}{c^2} \biggl ( \frac{L_\ell^2}{\omega^2} - 1 \biggr ) \exp \biggl [ \int_{0}^{r} \biggl ( \frac{N^2}{g} - \frac{g}{c^2} \biggr ) \mathrm{d}r \biggl ], \label{eq:rev4}
\end{eqnarray} 
in which we denote the local gravitational acceleration by $g$. 

\begin{figure}[t]
	\begin{center}
	\includegraphics[scale=0.36]{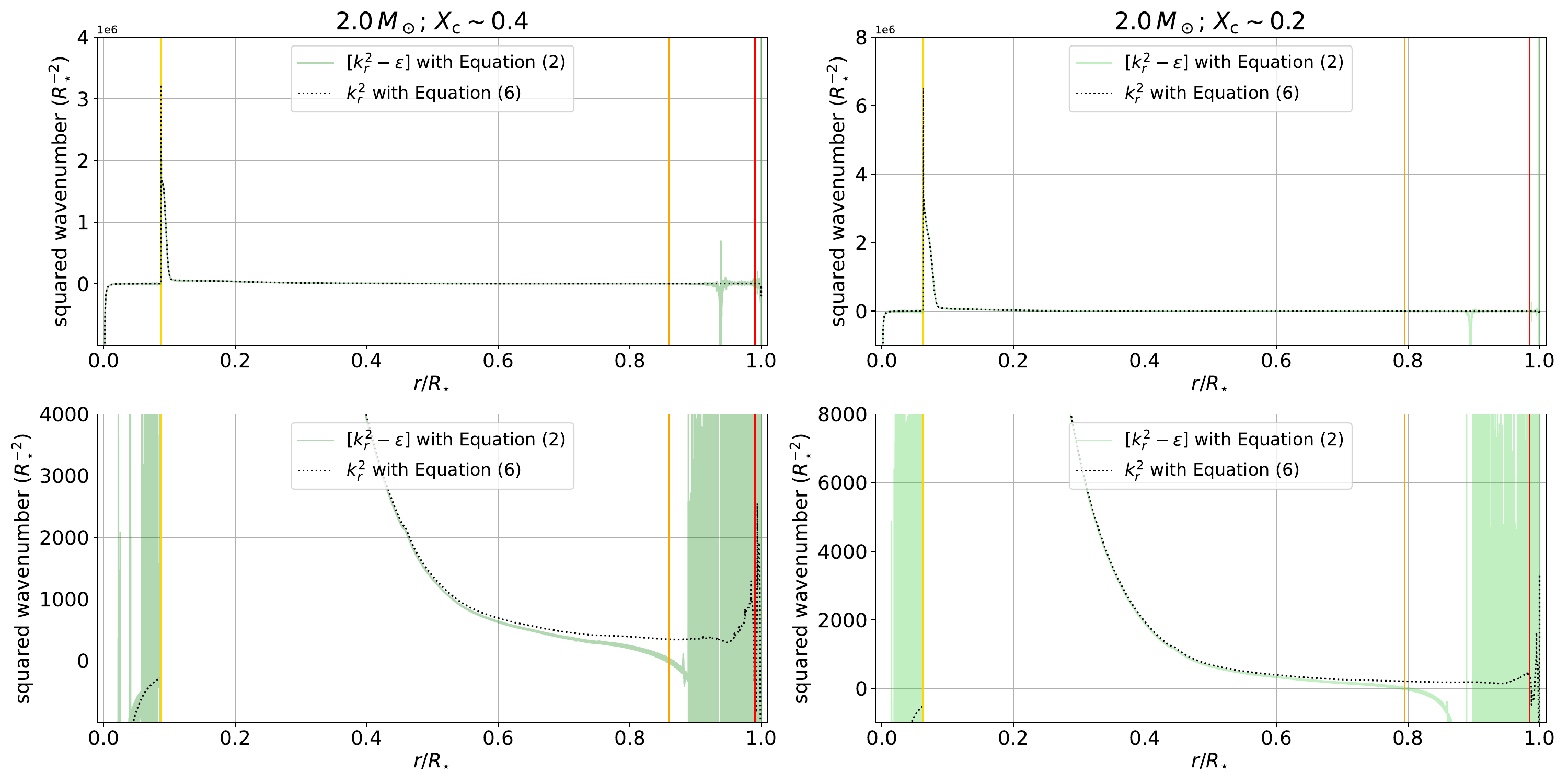}
		\caption{\footnotesize Comparison between $[k_r^2 - \epsilon]$ (the greenish curves, computed with Equation (\ref{eq:rev2})) and $k_r^2$ (the black dotted curves, computed with Equation (\ref{eq:2})) as a function of the fractional radius $r/R_\star$, where $R_\star$ is the stellar radius, in the case of $2 \, M_\odot$ main-sequence stellar models with different evolutionary stages, i.e., $X_\mathrm{c} = 0.4$ (left) and $0.2$ (right), where $X_\mathrm{c}$ is the hydrogen mass content at the center (details in the model computations can be found in Section \ref{sec:2-2} of the main text). 
		The eigenfrequency is chosen to be $1$ cycle per day, which is typical for the main-sequence g-mode pulsators \citep{2016MNRAS.455L..67M}. 
		It is apparent that $[k_r^2 - \epsilon]$ is well approximated by $k_r^2$ in almost all the region (see the top panels), although we find a few differences in the expanded looks (see the bottom panels). 
		While the locations of the lower turning point (yellow) are almost the same irrespective of whether $[k_r^2 - \epsilon]$ or $k_r^2$ is used, the locations of the upper turning point are slightly different (orange in the case of $[k_r^2 - \epsilon]$, and red in the case of $k_r^2$). 
		We also see the divergence of $[k_r^2 - \epsilon]$ in the outer envelope (see, e.g., $r/R_\star \sim 0.93$ in the left top panel), which is due to the fact that $P$ becomes zero at the point (see Equations (\ref{eq:rev3}) and (\ref{eq:rev4})). 
		Despite the differences above, it is expected that high-order g modes of intermediate-mass ($\sim 2 \, M_\odot$) main-sequence stars can be described by the approximated $k_r^2$ because we see little differences between $[k_r^2 - \epsilon]$ and $k_r^2$ in the deep region ($r /R_\star \sim 0.1$) where the radial wavenumber is so large that g-mode properties are mostly determined by the structure there. }
	\label{fig:rev1}
	\end{center}
\end{figure}

Equations (\ref{eq:rev1}) and (\ref{eq:rev2}) are exact under the Cowling approximation. 
To simplify the analysis of high-order g modes of intermediate-mass main-sequence stars, we then would like to proceed with the following approximations: 
\begin{eqnarray}
	\frac{\mathrm{d^2}v}{\mathrm{d}r^2} 
	+ 
	k_r^2 v = 0, \label{eq:1}
\end{eqnarray}
and 
\begin{eqnarray}
	k_r^2 = \frac{L^2 N^2}{\omega^2 r^2} - \frac{L^2}{r^2}. \label{eq:2}
\end{eqnarray} 
The validity of the approximations above can be mostly confirmed by comparing $[k_r^2 - \epsilon]$ where $k_r$ is given by Equation (\ref{eq:rev2}) and $k_r^2$ where $k_r$ is given by Equation (\ref{eq:2}). 
We show in Figure \ref{fig:rev1} the comparison in the case of $2 \, M_{\odot}$ main-sequence models (details in the model computation later given in Section \ref{sec:2-2}). 
The eigenfrequency 
is chosen to be $1$ cycle per day, which is typical for the main-sequence g-mode pulsators \citep{2016MNRAS.455L..67M}. 
It is apparent that in almost all the region $[k_r^2 - \epsilon]$ 
(greenish curves) is well approximated by $k_r^2$ 
(black dotted curves). 
(Note that the wiggles in $[k_r^2 - \epsilon]$ have numerical origins since we have to compute the second derivative of $P$.) 
Strictly speaking, the positions of the turning point, where the squared radial wavenumber changes the sign, are different. 
This is especially the case for the upper turning point; when we approximate the radial wavenumber by Equation (\ref{eq:2}), the upper turning point becomes shallower than that determined by where $[k_r^2 - \epsilon ] = 0$ (see the bottom panels of Figure \ref{fig:rev1} and compare the red vertical line and the orange vertical line). 
Nevertheless, the effects of the different locations of the upper turning point on high-order g-mode properties may be irrelevant because the high-order g-mode properties are mostly determined by the deep radiative region where $k_r^2$ is large (see the top panels of Figure \ref{fig:rev1} around $r/R_\odot \sim 0.1$). 
In other words, high-order g modes of intermediate-mass ($\sim 2 M_\odot$) main-sequence stars are nearly pure g modes, and thus, we may not need sophisticated isolation schemes as are suggested in the case of evolved stars \citep[see, e.g.,][for more discussions]{1977A&A....58...41A,2020ApJ...898..127O}. 
Based on the justification above, we will focus on analyzing Equations (\ref{eq:1}) and (\ref{eq:2}). 


As is evident in Equation (\ref{eq:2}), properties of high-order g modes are mostly determined by the BV frequency profile. 
In the case of intermediate-mass ($1.5 \, M_{\odot} < M < 3 \, M_\odot$) main-sequence stars, 
the BV frequency $N^2$ behaves as 
$N^2 < 0$ and $\omega^2 \gg |N^2| \sim 0$ in the convective core, and $N^2 \gg \omega^2 > 0$ in the radiative envelope. 
These features in the BV frequency profile result in the further approximations as follows: $k_r^2 \sim -L^2 / r^2 < 0$ in the convective core, and $k_r^2 \sim (L^2 N^2) / (\omega^2 r^2) > 0$ in the radiative envelope; therefore, g modes are evanescent (propagative) in the convective core (the radiative envelope). 
Throughout this paper, we persist on this simple description for the g-mode cavity of the intermediate-mass main-sequence stars, based on which we develop the perturbative analysis of high-order g modes. 
We also assume for simplicity that the outer edge of the radiative envelope is the surface of the star. 
As boundary conditions, we consider $v=0$ for the both center ($r=0$) and surface ($r=r_1$, where the radial coordinate of the stellar surface is denoted by $r_1$) of the star. 
Following U89, we then consider the Liouville transformation of Equation (\ref{eq:1}), i.e., $(r, v) \rightarrow (\zeta,W)$, leading to: 
\begin{eqnarray}
	\frac{\mathrm{d^2}W}{\mathrm{d}\zeta^2} 
	+ 
	\biggl [ 
	\zeta 
	- 
	f \biggl ( 
	\frac{\mathrm{d}r}{\mathrm{d}\zeta} \biggr )
	\biggr ] W 
	= 0,  \label{eq:3}
\end{eqnarray}
where the variable $\zeta$ is related to $r$ as: 
\begin{eqnarray}
	\zeta = 
	\mathrm{sgn}(k_r^2) 
	\biggl ( 
	\biggl | 
	\frac{3}{2} \int_{r_0}^{r} 
	|k_r| \mathrm{d}r
	\biggr |
	\biggr )^{2/3}, \label{eq:4}
\end{eqnarray}
and the variable $W$ is related to $v$ as: 
\begin{eqnarray}
	W = \biggl ( \biggl |  \frac{\mathrm{d}r}{\mathrm{d}\zeta} \biggr | \biggr )^{-1/2} v. \label{eq:4_1}
\end{eqnarray}
The bottom of the mode cavity, that is here assumed to be identical to the convective boundary, is denoted by $r_0$. 
The definition of the function $f$ is: 
\begin{eqnarray}
	f(q) 
	\equiv 
	|q|^{1/2} 
	\frac{\mathrm{d}^2 |q|^{-1/2}}{\mathrm{d} \zeta^2}, \label{eq:5}
\end{eqnarray}
for an arbitrary structural variable $q$. 
We will hereafter call it the $f$-term. 

We will apply the first-order perturbative analysis to solve Equation (\ref{eq:3}) in the following sections. 
Note that Equation (\ref{eq:3}) reduces to the Airy equation when we neglect the $f$-term, and that the eigenfunction is thus given as the Airy function under suitable boundary conditions. 
The Airy functions can be further approximated as the JWKB solutions well inside the mode cavity (more details can found in Chapter 16 of U89). 

\subsection{The $f$-term} \label{sec:2-2}
As mentioned in the end of the last section, one of the keys to the asymptotic analysis of high-order modes is the assumption that the $f$-term is negligible compared with $\zeta$. 
In this section, we would like to show how the $f$-term behaves in stellar interiors. 
To this end, 
1-d main-sequence stellar models with different masses ($M = 2$ and $4 \, M_\odot$) and evolutionary stages ($X_{\mathrm{c}} \sim 0.4$ and $0.2$, where $X_{\mathrm{c}}$ is the hydrogen mass fraction at the center) are computed via MESA \citep[version 15140;][]{2011ApJS..192....3P,2013ApJS..208....4P,2015ApJS..220...15P,2018ApJS..234...34P,2019ApJS..243...10P}. 
The initial chemical composition is $(X,\, Y, \, Z) = (0.70, \, 0.28, \, 0.02)$. 
We have adopted default settings in MESA except that we have activated elemental diffusion \citep[see, e.g.,][]{2018ApJS..234...34P} in order to eliminate discontinuities in the BV frequency profiles that are suspected to be unphysical. 
The elemental diffusion is also weakened by setting $\mathtt{diffusion\_v\_max}$ to be $10^{-7}$ to avoid complete depletion of the helium in the envelope \citep{2002A&A...390..611M}. 
It should also be noted that overshooting beyond the convective boundary is not taken into account in the models. 

\begin{figure}[t]
	\begin{center}
	\includegraphics[scale=0.34]{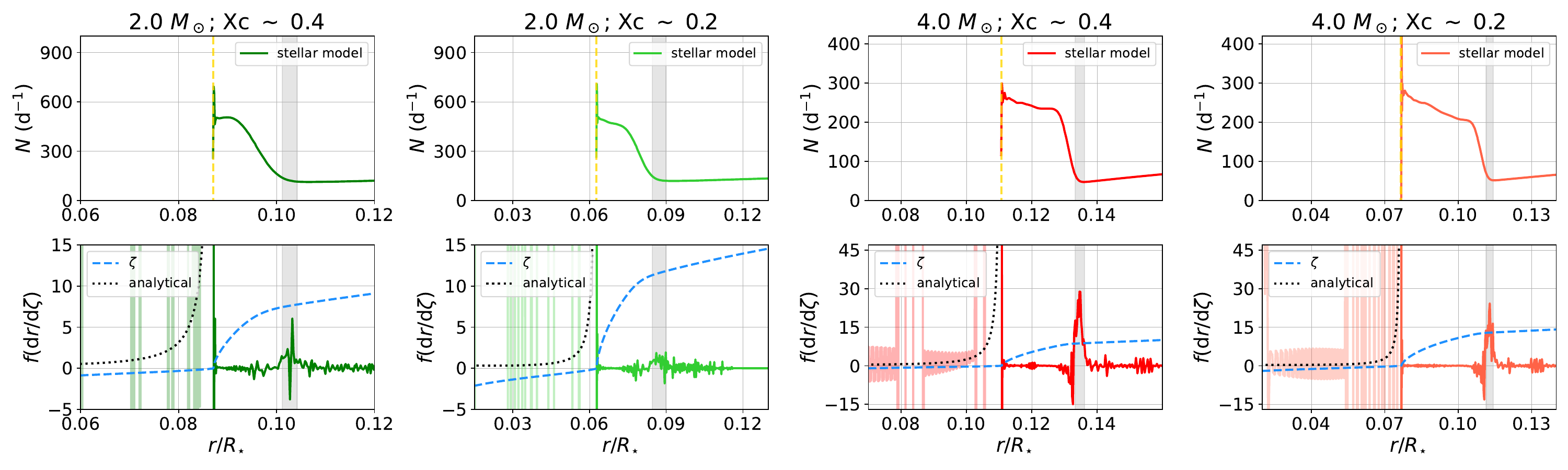}
		\caption{\footnotesize Internal structures around the convective core boundary of the four stellar models with different masses ($M = 2 \, M_\odot$ or $4 \, M_\odot$, greenish or reddish, respectively) and different evolutionary stages ($X_\mathrm{c} \sim 0.4$ or $0.2$, darker or lighter, respectively). 
		The top row shows the BV frequency profiles $N$ in units of $\mathrm{d}^{-1}$ as a function of the stellar fractional radius $r/R_\star$, where $R_\star$ is the stellar radius. 
		The convective boundary is indicated by the yellow dashed line below which there is no value for $N$ because $N^2<0$ in the convective core. 
		Just above the convective core boundary is the hump in $N$, which originates from the chemical composition gradients left behind the receding the nuclear burning core \citep[e.g.,][]{2008MNRAS.386.1487M}. 
		The region where the first derivative of the BV frequency transitions sharply is highlighted by the grey shaded area. 
		The bottom row shows the $f$-terms (solid curves) of the corresponding stellar models in the same ranges of the fractional radius as are shown in the top row. 
		Because numerical differentiation of $N$ of the stellar models leads to the shaky $f$-terms especially inside the convective core, we lighten the color of the $f$-term in the convective core for visual aid. 
		We also draw analytical $f$-terms (black dotted) in the convective core that is computed with Equation (\ref{eq:5}) assuming $k_r^2 \sim -L^2 / r^2$. 		
		The blue dashed curve represents $\zeta$ that is defined as Equation (\ref{eq:7}). 
		It is seen that, in the radiative zone where $N$ is finite, $\zeta$ is much larger than the $f$-terms except for the grey shaded area where the first derivative of $N$ transitions sharply. 
		In contrast, inside the convective core, the $f$-term (black dotted) is comparable to or larger than (the absolute value of) $\zeta$ (blue dashed). 
		}
	\label{fig:1}
	\end{center}
\end{figure}

The bottom panels of Figure \ref{fig:1} show comparisons between the $f$-term and $\zeta$ for the four stellar models. 
It is evident that the $f$-term is not negligible everywhere. 
In the following paragraphs, we give a few comments on each region where the $f$-term is non-negligible, namely, the convective core, the convective boundary, and a part of the radiative zone where the first derivative of the BV frequency varies sharply. 

Inside the convective core, the $f$-term (the black dotted curves in the bottom panels of Figure \ref{fig:1}) is mostly comparable to or larger than (the absolute value of) $\zeta$ (the blue dashed curves), and thus, the solution of Equation (\ref{eq:3}) cannot be described by the Airy function there. 
Nevertheless, because 
the radial wavenumber can be approximated by $k_r^2 \sim -L^2 / r^2$ (see discussions in Section \ref{sec:2-1}), we can relatively easily solve Equation (\ref{eq:1}). 
We will revisit this point later in Section \ref{sec:2-3}. 

Near the convective boundary, the $f$-term is quite large due to the almost discontinuous change in the BV frequency profile. 
In addition, $\zeta$ is by definition close to zero in the vicinity of the convective boundary (see Equation (\ref{eq:4})). 
Therefore, similarly to the case of the convective core, Equation (\ref{eq:3}) cannot be approximated by the Airy equation, and we cannot describe the solution with the Airy function in this region. 


The situation is different in the radiative zone. 
We see that the $f$-term is mostly much smaller than $\zeta$, which guarantees us to approximate Equation (\ref{eq:3}) as the Airy equation. 
One exception is the region where the first derivative of the BV frequency sharply transitions (the grey shaded regions in the top and bottom panels of Figure \ref{fig:1}). 
It is generally considered that such a sharp transition in the BV frequency is related to the steep chemical composition gradient left behind by the receding nuclear burning core along stellar evolution \citep[e.g.,][]{2008MNRAS.386.1487M}.  

For the radiative zone, the ratio of the $f$-term to $\zeta$ around the sharp transition in the BV frequency depends on the mass of the stellar models. 
In the case of $2 \, M_{\odot}$ models, the ratio is $\sim$ $1-10$ $\%$, thus hinting at the possibility that the $f$-term can be regarded as a small perturbation. 
On the other hand, in the case of $4 \, M_{\odot}$ models, the $f$-term can be comparable to or even higher than $\zeta$, and thus, the first-order perturbative approach cannot be applicable obviously. 
It should nevertheless be instructive to mention that H23's semi-analytical expression can reproduce $\Delta P_g$ patterns of the higher-mass ($> 3 \, M_{\odot}$) models (see Figure 10 in H23). 
This is because the variations in the BV frequencies of the relatively high-mass models are sharp enough for us to treat it as a discontinuity in the first derivative of the BV frequency, which allows us to utilize H23's semi-analytical approach (see H23 for more discussions.) 
We will thus not consider applying the perturbation theory to the models with the mass higher than $3 \, M_{\odot}$, and will from now on focus on the models with the mass lower than $3 \, M_{\odot}$. 

\subsection{Unperturbed equation} \label{sec:2-3}
We then move on to the perturbative analysis of Equation (\ref{eq:3}). 
The first step is to choose an unperturbed equation. 
A naive choice may be the Airy equation as the unperturbed state, but this choice would be inappropriate since the $f$-term can be comparable to or larger than $\zeta$ in some parts of stars (see Figure \ref{fig:1}); the $f$-term as a whole is too large for us to treat it as a small perturbation. 
Therefore, we take another equation as the unperturbed state as is described in the following paragraphs. 

An important point is that, regarding oscillatory components in the $\Delta P_g$ patterns, 
only the 
sharp transition in the BV frequency well inside the g-mode cavity is relevant. 
This is because gravity waves are propagative inside the radiative regions; the structural variations in the convective core cannot affect the mode trapping of g modes. 
The structural variations around the convective boundary may be included in the bottom part of the radiative zone, and they can affect the mode trapping. 
Still, the effect on the $\Delta P_g$ pattern appears as an oscillatory component with quite a long period ($> 100$ in the radial order $n$) because the structural variations are located fairly close to the convective boundary (see also discussions in Appendix A of H23). 
Thus, to first approximation, such a long-period oscillatory component can be regarded as just a linear trend in the $\Delta P_g$ pattern. 

Keeping the discussions above in mind, we divide the $f$-term into two components: the $f_0$-term that is related to structural variations in the convective core and those around the convective boundary, and the $f_1$-term that is related to a sharp transition in the BV frequency well inside the g-mode cavity, i.e., $f = f_0 + f_{1}$. 
In other words, the $f_0$-term is the $f$-term of a stellar model that has such a smooth chemical composition profile that the second-derivative of the BV frequency (see Equation (\ref{eq:6})) is zero well inside the radiative zone. 
The separation above leads to the following unperturbed equation: 
\begin{eqnarray}
	\frac{\mathrm{d^2}W}{\mathrm{d}\zeta^2} 
	+ 
	\biggl [\zeta  - f_0 \biggl ( \frac{\mathrm{d}r}{\mathrm{d}\zeta} \biggr )\biggr ] W
	= 0. \label{eq:6}
\end{eqnarray} 

Note that the unperturbed equation takes slightly different forms for the radiative envelope and the convective core because the radial wavenumber behaves differently in each region. 
As we discuss in Section \ref{sec:2-1}, in the case of high-order g modes of ordinary 1-d stellar models, $N^2 \gg \omega^2$ in the radiative envelope, and the radial wavenumber can be approximated as $k_r^2 \sim (L^2 N^2) / (\omega^2 r^2)$, meanwhile $\omega^2 \gg |N^2| \sim 0$ in the convective core, and $k_r^2 \sim -L^2 / r^2$. 
Consequently, the frequency dependence of $\zeta$ is different for the radiative and convective zones as below: 
\begin{equation}
  \zeta = 
  \begin{dcases}
    - \biggl ( \frac{3}{2} \int_{r}^{r_0} \frac{L}{r'} \mathrm{d}r' \biggr ) ^{2/3} & \text{($r < r_0$; convective core)} \\
    0                 & \text{($r=r_0$; convective boundary)} \\
    + \biggl ( \frac{3}{2} \int_{r_0}^{r} \frac{LN}{\omega_n r'} \mathrm{d}r' \biggr ) ^{2/3} 
    = \omega_n^{-2/3} \biggl ( \frac{3}{2} \int_{r_0}^{r} \frac{LN}{r'} \mathrm{d}r' \biggr ) ^{2/3} 
    = \omega_n^{-2/3} \biggl ( \frac{3}{2} \biggr )^{2/3} \zeta' & \text{($r > r_0$; radiative envelope)}
  \end{dcases} \label{eq:7}
\end{equation} 
In the last line of the equation above, the radial order $n$ is explicitly shown for the eigenfrequency (i.e. $\omega_n$), 
and we have also introduced a new coordinate $\zeta'$ to render the frequency dependence in $\zeta$ to be explicit, namely, 
\begin{eqnarray}
	\zeta' = \biggl [ \int_{r_0}^{r} \frac{LN}{r'} \mathrm{d}r' \biggr ]^{2/3}. \label{eq:8}
\end{eqnarray} 

One remarkable point about Equation (\ref{eq:7}) is an absence of the frequency dependence in the unperturbed equation for the convective core (see the first line of Equation (\ref{eq:7})); the solution $W$ is independent of the eigenfrequency. 
This can be confirmed from a different perspective by taking a look at Equation (\ref{eq:1}). 
Since the radial wavenumber can be approximated as $k_r^2 \sim -L^2 / r^2$ in the convective core, we can actually derive $v \propto r^{\ell + 1}$ that applies for all the g modes; all the g-mode eigenfunctions (with the same spherical degree $\ell$) behave in the same manner in the convective core. 
Therefore, for a fixed spherical degree $\ell$, we may set $v = (r_{0})^{\ell+1}$ at the convective boundary ($r = r_0$) as the inner boundary condition instead of $v = 0$ at $r =0$. 
This resetting of the inner boundary condition in short means that we do not need to care about the convective core when we apply the perturbation theory to describe the oscillatory component in the $\Delta P_g$ pattern. 

\subsection{Variational principle} \label{sec:2-4}
Based on the unperturbed equation (Equation (\ref{eq:6})), we can conduct the perturbative analysis to establish a relationship among the $f_1$-term, the perturbed eigenfrequency, and the perturbed eigenfunction. 
Importantly, to the first-order, we can neglect effects of the perturbed eigenfunction on the eigenfrequency; the eigenfrequency is stationary to the first-order with respect to perturbations in the eigenfunction (variational principle). 
Before deriving a linear relation between the $f_1$-term and the $\Delta P_g$ patterns, we here would like to show that the variational principle holds. 

Firstly, we consider perturbations in the eigenfrequency ($\Delta \omega_n$) and the corresponding eigenfunction ($\Delta W_n$) to the unperturbed equation (Equation (\ref{eq:6})). 
For the radiative zone, the relationship between $\Delta \omega_n$ and $\Delta W_n$ is expressed by  the following equation: 
\begin{eqnarray}
	\frac{\mathrm{d^2}}{\mathrm{d}\zeta'^2} (W_n + \Delta W_n)
	+ 
	\biggl [(\omega_n + \Delta \omega_n)^{-2} \biggl ( \frac{3}{2} \biggr )^2 \zeta' - f_0 \biggl ( \frac{\mathrm{d}r}{\mathrm{d}\zeta'} \biggr )\biggr ] (W_n + \Delta W_n)
	= 0, \label{eq:9}
\end{eqnarray}
where we explicitly show the radial order $n$ for the eigenfunction (i.e. $W_n$). 
It should be noticed that the frequency-independent coordinate $\zeta'$ (Equation (\ref{eq:8})) is used instead of $\zeta$. 
Retaining the first-order terms in Equation (\ref{eq:9}) ends up with: 
\begin{eqnarray}
	-\biggl ( \frac{3}{2} \biggr )^{2} (-2 \omega_n^{-3}) \zeta' W_n \Delta \omega_n 
	= 
	\frac{\mathrm{d}^2 \Delta W_n}{\mathrm{d} \zeta'^2} + \biggl [ \omega_n^{-2} \biggl ( \frac{3}{2} \biggr )^2 \zeta' - f_0 \biggl ( \frac{\mathrm{d}r}{\mathrm{d}\zeta'} \biggr ) \biggr ] \Delta W_n. \label{eq:10}
\end{eqnarray}

We then take the inner product for the both sides of Equation (\ref{eq:10}) with the eigenfunction $W_n$ over the coordinate $\zeta'$. 
As we reformulated the eigenvalue problem so that the inner and outer boundary conditions are given at the both edges of the radiative zone (see Section \ref{sec:2-3}), the integration domain is $0 \le \zeta' \le \zeta'_1$, 
thus leading to the following equation: 
\begin{eqnarray}
	\int_{0}^{\zeta'_1} \biggl ( \frac{3}{2} \biggr )^{2} (2 \omega_n^{-3}) \zeta' |W_n|^2 \mathrm{d} \zeta' \times \Delta \omega_n 
	= 
	\int_{0}^{\zeta'_1} \biggl \{ \frac{\mathrm{d}^2 \Delta W_n}{\mathrm{d} \zeta'^2} + \biggl [ \omega_n^{-2} \biggl ( \frac{3}{2} \biggr )^2 \zeta' - f_0 \biggl ( \frac{\mathrm{d}r}{\mathrm{d}\zeta'} \biggr ) \biggr ] \Delta W_n \biggr \} W_n \mathrm{d}\zeta', \label{eq:11}
\end{eqnarray}
where the value of $\zeta'$ at the outer edge of the g-mode cavity 
is represented by $\zeta'_1$. 

The integration in the right-hand side of Equation (\ref{eq:11}) can be computed in the following ways: 
\begin{eqnarray} 
& \, & 
	\int_{0}^{\zeta'_1} \biggl \{ \frac{\mathrm{d}^2 \Delta W_n}{\mathrm{d} \zeta'^2} 
	+ \biggl [ \omega_n^{-2} \biggl ( \frac{3}{2} \biggr )^2 \zeta' - f_0 \biggl ( \frac{\mathrm{d}r}{\mathrm{d}\zeta'} \biggr ) \biggr ] \Delta W_n \biggr \} W_n \mathrm{d}\zeta' \nonumber \\
& = & 
	\biggl [ \frac{\mathrm{d} \Delta W_n}{\mathrm{d} \zeta'} W_n \biggr ]_{0}^{\zeta'_1} 
	- \int_{0}^{\zeta'_1} \frac{\mathrm{d} \Delta W_n}{\mathrm{d} \zeta'} \frac{\mathrm{d} W_n}{\mathrm{d} \zeta'} \mathrm{d} \zeta' 
	+ \int_{0}^{\zeta'_1} \biggl [ \omega_n^{-2} \biggl ( \frac{3}{2} \biggr )^2 \zeta' - f_0 \biggl ( \frac{\mathrm{d}r}{\mathrm{d}\zeta'} \biggr ) \biggr ] (\Delta W_n)  W_n \mathrm{d}\zeta' \nonumber \\
& = & 
	\biggl [ \frac{\mathrm{d} \Delta W_n}{\mathrm{d} \zeta'} W_n \biggr ]_{0}^{\zeta'_1} 
	- \biggl [ \Delta W_n \frac{\mathrm{d} W_n}{\mathrm{d} \zeta'}  \biggr]_{0}^{\zeta'_1} 
	+ \int_{0}^{\zeta'_1} \Delta W_n \frac{\mathrm{d^2} W_n}{\mathrm{d} \zeta'^2} \mathrm{d} \zeta' 
	+ \int_{0}^{\zeta'_1} \biggl [ \omega_n^{-2} \biggl ( \frac{3}{2} \biggr )^2 \zeta' - f_0 \biggl ( \frac{\mathrm{d}r}{\mathrm{d}\zeta'} \biggr ) \biggr ] (\Delta W_n)  W_n \mathrm{d}\zeta' \nonumber \\ 
& = & 
	\biggl [ \frac{\mathrm{d} \Delta W_n}{\mathrm{d} \zeta'} W_n \biggr ]_{0}^{\zeta'_1} 
	- \biggl [ \Delta W_n \frac{\mathrm{d} W_n}{\mathrm{d} \zeta'}  \biggr]_{0}^{\zeta'_1} 
	+ \int_{0}^{\zeta'_1} \biggl [\frac{\mathrm{d^2} W_n}{\mathrm{d} \zeta'^2} 
	+ \biggl \{ \omega_n^{-2} \biggl ( \frac{3}{2} \biggr )^2 \zeta' - f_0 \biggl ( \frac{\mathrm{d}r}{\mathrm{d}\zeta'} \biggr ) \biggr \} W_n \biggr ] \Delta W_n  \mathrm{d}\zeta' \nonumber \\ 
& = & 0. \label{eq:12}
\end{eqnarray} 
The same boundary conditions apply for the perturbed eigenfunctions, e.g., $W_n = 0$ and $\Delta W_n = 0$ at the outer edge of the mode cavity ($\zeta' = \zeta'_1$). 
In addition, as we discussed in Section \ref{sec:2-3}, g-mode eigenfunctions are the same in the convective core for a fixed spherical degree, and therefore, $\Delta W_n = 0$ and $\mathrm{d} \Delta W_n / \mathrm{d} \zeta = 0$ at $\zeta = 0$. 
The surface integrals in Equation (\ref{eq:12}) consequently are zero. 
The third term in the fourth line of Equation (\ref{eq:12}) is zero because of Equation (\ref{eq:6}). 
As a result, the perturbed eigenfrequency $\Delta \omega_n = 0$ (see Equation (\ref{eq:11})); the variational principle holds in this case. 


\subsection{Perturbed equation} \label{sec:2-5}
In this section, we derive a linear relationship between the $f_1$-term and the perturbed eigenfrequency $\Delta \omega_n$. 
As we discussed in the previous sections, all we need to care is the radiative zone, and the perturbed equation is: 
\begin{eqnarray}
	\frac{\mathrm{d^2} W_n}{\mathrm{d}\zeta'^2} 
	+ 
	\biggl [(\omega_n + \Delta \omega_n)^{-2} \biggl ( \frac{3}{2} \biggr )^2 \zeta' 
		- f_0 \biggl ( \frac{\mathrm{d}r}{\mathrm{d}\zeta'} \biggr ) 
		- f_1 \biggl ( \frac{\mathrm{d}r}{\mathrm{d}\zeta'} \biggr ) \biggr ] W_n
	= 0, \label{eq:13}
\end{eqnarray}
where the perturbed eigenfunction $\Delta W_n$ has been omitted due to the variational principle (Section \ref{sec:2-4}). 

In a manner similar to what we demonstrated in the last section, we linearize terms in the equation above, and we then retain the first-order terms, namely, $\Delta \omega_n$ and the $f_1$-term, which leads to the following equation: 
\begin{eqnarray}
	-2 \omega_n^{-3} \biggl ( \frac{3}{2} \biggr )^{2} \zeta' W_n \times \Delta \omega_n
	- f_1 W_n = 0. \label{eq:14}
\end{eqnarray}
By taking the inner product for the both sides of Equation (\ref{eq:14}) with $W_n$ in the range $0 \le \zeta' \le \zeta'_1$, we have a linear relationship between the perturbed eigenfrequency $\Delta \omega_n$ and the $f_1$-term as below: 
\begin{eqnarray}
	\Delta \omega_n = 
	- \frac{\omega_n^3}{2} \biggl ( \frac{3}{2} \biggr )^{-2}  
		\frac{\int_{0}^{\zeta'_1} f_1 |W_n|^2 \mathrm{d} \zeta'}{\int_{0}^{\zeta'_1} \zeta' |W_n|^2 \mathrm{d} \zeta'} 
	= \int_0^{\zeta'_1} K_{n}(\zeta') f_1(\zeta') \mathrm{d} \zeta', \label{eq:15}
\end{eqnarray}
where in the last line we have introduced the sensitivity kernel $K_n(\zeta')$ that is defined as: 
\begin{eqnarray}
	K_n(\zeta') = 
	- \frac{\omega_n^3}{2} \biggl ( \frac{3}{2} \biggr )^{-2}  
		\frac{|W_n|^2}{\int_{0}^{\zeta'_1} \zeta' |W_n|^2 \mathrm{d} \zeta'}. \label{eq:16}
\end{eqnarray}
Thus, the perturbed eigenfrequency is regarded as a weighted average of the $f_1$-term in which the kernel $K_n(\zeta')$ acts as a weighting function. 

By definition, $\omega_n = 2 \pi / P_n$, so that $\Delta \omega_n = - 2 \pi P_{n}^{-2} \Delta P_n$ to the first order, based on which we can relate the perturbed eigenperiod $\Delta P_n$ to the $f_1$-term. 
It should be noted that $\Delta P_n$ is not the period spacing ($\Delta P_g = P_{n+1} - P_n$), but it is the difference between the eigenperiod of the perturbed state ($P_{n,\mathrm{ptb}}$) where the $f_1$-term is taken into account and that of the unperturbed state ($P_{n,\mathrm{unptb}}$). 
Accordingly, the period spacing $\Delta P_g$ is computed in the following way: 
\begin{eqnarray}
	\Delta P_g \equiv P_{n+1,\mathrm{ptb}} - P_{n,\mathrm{ptb}} 
	= (P_{n+1,\mathrm{unptb}} + \Delta P_{n+1}) - (P_{n,\mathrm{unptb}} + \Delta P_{n}) 
	= \Delta P_{g,\mathrm{unptb}} + (\Delta P_{n+1} - \Delta P_{n}), \label{eq:17}
\end{eqnarray}
where the period spacing for the unperturbed state $\Delta P_{g,\mathrm{unptb}}$ is expressed as (see U89): 
\begin{eqnarray}
	\Delta P_{g,\mathrm{unptb}} = \frac{2 \pi^2}{L} \biggl [ \int_{r_0}^{r_1} \frac{N}{r} \mathrm{d}r \biggr ]^{-1}. \label{eq:18}
\end{eqnarray}
We will see if the first-order perturbative formulations presented in this section would be useful or not for explaining $\Delta P_g$ patterns (Sections \ref{sec:3} and \ref{sec:4}). 

\section{Validation tests with toy models of the BV frequency} \label{sec:3}
In this section, with some toy models of the BV frequency profiles, we carry out validation tests of the analytical expression that is derived based on the first-order perturbation theory. 
We firstly present how we construct toy models of the BV frequency profiles (Section \ref{sec:3-1}). 
After we describe the kernels that are required for evaluating the integration in Equation (\ref{eq:15}) (Section \ref{sec:3-2}), we compare $\Delta P_g$ patterns numerically computed 
and those computed based on the first-order perturbative analysis (Section \ref{sec:3-3}). 

\subsection{Toy models of the BV frequency} \label{sec:3-1}
To render toy models to represent the BV frequency profile of stars with masses $\sim 2 \, M_{\odot}$ as nicely as possible, we firstly compute a series of stellar models via MESA, based on whose BV frequency profiles we construct the toy models. 
Using the same settings for MESA as described in Section \ref{sec:2-2}, we computed stellar models with different masses ($M = 1.6$ and $2 \, M_{\odot}$) and evolutionary stages ($X_{\mathrm{c}} \sim 0.5$, $0.4$, $0.3$, and $0.2$). 

We subsequently extract information on the sharp transition in the BV frequencies of the stellar models. 
Specifically, we fit the hyperbolic tangent function to the first derivative of $(N/r)^{-1/2}$. 
The functional form of the hyperbolic function is: 
\begin{eqnarray}
	H(r) = -A_{N} \tanh \biggl ( \frac{r- r_\star}{\sigma_r} \biggr ) + h, \label{eq:19}
\end{eqnarray}
where the amplitude, the width, and the central location of the sharp transition in the first derivative of $(N/r)^{-1/2}$ are denoted by $A_N$, $\sigma_r$, and $r_\star$, respectively. 
We assumed that $h=A_N+h_\mathrm{c}$ to force $H(r)$ to be positive, in which $h_\mathrm{c}$ is a positive constant. 
The choice of the value of $h_\mathrm{c}$ does not have a significant impact on the toy BV frequencies as long as $h_\mathrm{c} \ll A_N$. 
We have adopted $h_\mathrm{c} = 0.1$ in this study. 
The fitting is conducted for a region where the $f_1$-term is non-negligible compared with $\zeta$ (see the grey shaded areas in Figure \ref{fig:1}). 
The reason why we focus on the first derivative of $(N/r)^{-1/2}$ is that the $f$-term is essentially the second derivative of $(N/r)^{-1/2}$ (see, e.g., Equation (\ref{eq:5})). 
We give more discussions on the relationship between the $f$-term and $(N/r)^{-1/2}$ in the second paper of this series (paper 2). 

\begin{figure}[t]
	\begin{center}
	\includegraphics[scale=0.34]{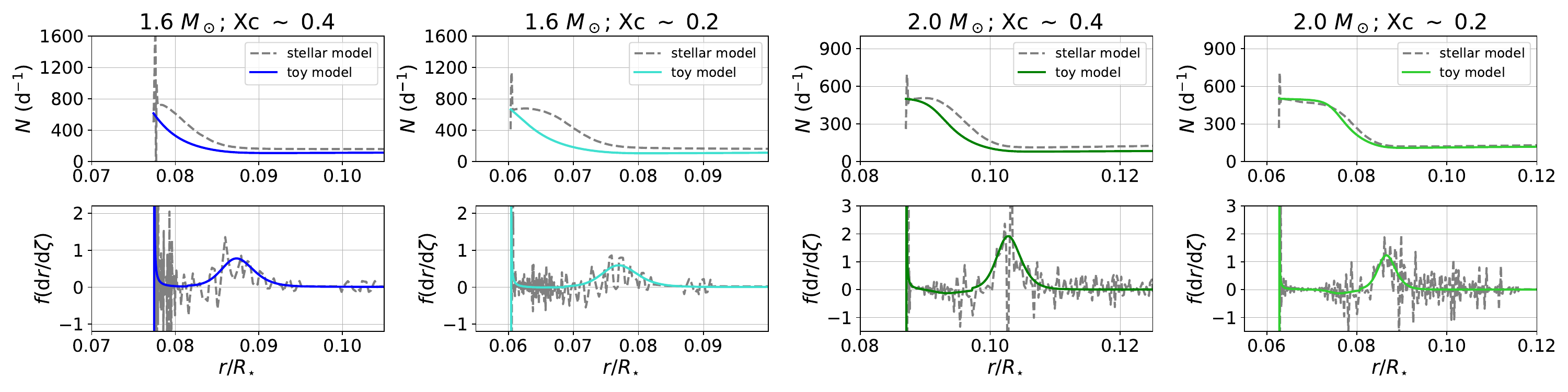}
		\caption{\footnotesize Comparisons between the toy models (colored curves) and the stellar models (grey dashed curves) with different masses ($M = 1.6 \, M_\odot$ or $2 \, M_\odot$, blueish or greenish, respectively) and different evolutionary stages ($X_\mathrm{c} \sim 0.4$ or $0.2$, darker or lighter, respectively). 
		In the same manner as in Figure \ref{fig:1}, the BV frequencies $N$ (in units of $\mathrm{d}^{-1}$) and the $f$-terms are shown in the top and bottom rows, respectively. 
		We see good agreement between the toy models and stellar models, which is especially the case for the $f$-terms. 
		Note that $N$ and the $f$-term are shown only for the radiative zone because the convective core is not relevant in the first-order perturbative approach we demonstrated in this study (see discussions in Section \ref{sec:2-3}). 
	}
	\label{fig:2}
	\end{center}
\end{figure}

Finally, the fitted hyperbolic tangent model is integrated to compute $(N/r)^{-1/2}$ with which we obtain $N$. 
The integration is carried out starting from the convective boundary toward the surface. 
The initial value for $(N/r)^{-1/2}$ at the convective boundary is given by that of the corresponding stellar model. 
The top panels of Figure \ref{fig:2} show comparisons between the BV frequency profiles of the stellar models (the grey dashed curves) and the toy models thus constructed (the colored curves.) 
It is seen that the toy models can more or less reproduce the smooth transitions in the BV frequencies. 
In particular, the toy models can very well reproduce the $f_1$-terms of the stellar BV frequencies (see the bottom panels of Figure \ref{fig:2}). 

\subsection{Kernels} \label{sec:3-2}
Before validation tests of the analytical expression of the $\Delta P_g$ pattern, we would like to specify the explicit expression of the eigenfunction $W_n$ that is required for computing the kernel $K_{n}$ (see Equation (\ref{eq:16})). 
As we discuss in Sections \ref{sec:2-2} and \ref{sec:2-3}, in the unperturbed state where we do not consider the $f_1$-term, we can approximate Equation (\ref{eq:3}) as the Airy equation well inside the radiative zone. 
Although it is customary for us to describe $W_n$ only by the first kind of the Airy function to force $W_n$ to be regular at the center (see U89), we can relax this condition because the regularity of the eigenfunction at the center is already verified via the functional form of $v$, i.e., $v \propto r^{\ell + 1}$ (see Section \ref{sec:2-3}). 
Thus, the eigenfunction $W_n$ can be expressed as $W_n(\zeta) = a \mathrm{Ai}(\zeta) + b \mathrm{Bi}(\zeta)$, where $\mathrm{Ai}(\zeta)$ and $\mathrm{Bi}(\zeta)$ are the first and second kind of the Airy functions, respectively, and $a$ and $b$ are coefficients of the linear combination. 
All we need for specifying $W_n$ is to determine the coefficients $a$ and $b$. 

It should be noticed that we can describe $W_n$ by $a \mathrm{Ai}(\zeta) + b \mathrm{Bi}(\zeta)$ almost everywhere in the radiative zone except for the very narrow region just above the convective core (see the bottom panels in Figure \ref{fig:1}, and compare the $f$-term and $\zeta$ around the convective boundary); to first approximation, let us then describe $W_n$ by $a \mathrm{Ai}(\zeta) + b \mathrm{Bi}(\zeta)$ for \textit{all} the radiative zone down to the convective boundary. 
This in essence is identical to that we neglect the structural variations around the convective boundary. 
Remembering that $v$ and $W_n$ are related to each other as $v = (|\mathrm{d}r / \mathrm{d}\zeta|)^{1/2} W_n$ (see Equation (\ref{eq:4_1})), admitting $W_n = a \mathrm{Ai}(\zeta) + b \mathrm{Bi}(\zeta)$ for the whole radiative zone, despite such a crude approximation, allows us to consider the boundary conditions at the convective boundary ($r = r_0$) as below: 
\begin{eqnarray}
	v(r = r_0) = \biggl [ \biggl (  \frac{\mathrm{d}r}{\mathrm{d}\zeta} \biggr )^{1/2} [ a \mathrm{Ai}(\zeta) + b \mathrm{Bi}(\zeta)] \biggr ] \biggr |_{r_0}, \label{eq:20}
\end{eqnarray}
and 
\begin{eqnarray}
	\frac{\mathrm{d}v}{\mathrm{d} r}\biggl |_{r_0} 
	= 
	\frac{\mathrm{d}}{\mathrm{d}r} \biggl [ \biggl (  \frac{\mathrm{d}r}{\mathrm{d}\zeta} \biggr )^{1/2} [ a \mathrm{Ai}(\zeta) + b \mathrm{Bi}(\zeta)] \biggr ] \biggr |_{r_0}. \label{eq:21}
\end{eqnarray}

We then consider the behavior of $\mathrm{d}\zeta / \mathrm{d}r$ around the convective boundary ($r = r_0$). 
As we noted in Section \ref{sec:2-2}, we are not taking into account the convective overshooting. 
In this case, the outer edge of the uniformly mixed core region is identical to the Schwarzschild boundary where the BV frequency transitions from negative ($N^2 \sim 0$ in the convective region) to positive ($N^2 \gg \omega^2$ in the radiative region); 
the BV frequency transitions almost discontinuously from negative to positive so that $\mathrm{d}\zeta / \mathrm{d}r \rightarrow + \infty$ at $r = r_0$. 
(Note that the behavior of $\mathrm{d}\zeta / \mathrm{d}r$ is different when there is an overshoot zone; the Schwarzschild boundary, at which the BV frequency changes the sign, can be inside the uniformly mixed core region, and the BV frequency can vary smoothly just above the Schwarzschild boundary, possibly rendering $\mathrm{d}\zeta / \mathrm{d}r$ to be finite at the outer edge of the overshoot zone; see, e.g., \citet{2022ApJ...930...94P} for more discussions on the BV frequency profile in the overshooting region.) 

By assuming that $\mathrm{d}\zeta / \mathrm{d}r$ at $r = r_0$ can be expressed by a ``huge'' finite number $c$ and assuming that $\mathrm{d}\zeta / \mathrm{d}r$ varies proportionally to $r$ in the vicinity of $r=r_0$ so that the first derivative of $\mathrm{d}\zeta / \mathrm{d}r$ (in terms of $r$) is negligible, the boundary conditions can be rewritten as: 
\begin{eqnarray}
	v(r = r_0) = c^{-1/2} [ a \mathrm{Ai}(\zeta = 0) + b \mathrm{Bi}(\zeta = 0)], \label{eq:22}
\end{eqnarray}
and 
\begin{eqnarray}
	\frac{\mathrm{d}v}{\mathrm{d} r}\biggl |_{r_0} 
	= 
	 \biggl (  \frac{\mathrm{d}r}{\mathrm{d}\zeta} \biggr |_{r_0} \biggr )^{1/2} \biggl [ a \frac{\mathrm{d}\mathrm{Ai}}{\mathrm{d}\zeta} \frac{\mathrm{d}\zeta}{\mathrm{d}r} +  b \frac{\mathrm{d}\mathrm{Bi}}{\mathrm{d}\zeta} \frac{\mathrm{d}\zeta}{\mathrm{d}r} \biggr ] \biggr |_{r_0}
	 = 
	 c^{1/2} \biggl [ a \frac{\mathrm{d}\mathrm{Ai}}{\mathrm{d}\zeta} \biggl |_{\zeta=0} +  b \frac{\mathrm{d}\mathrm{Bi}}{\mathrm{d}\zeta} \biggl |_{\zeta=0} \biggr ]. \label{eq:23}
\end{eqnarray}

Solving the system of the equations (Equations (\ref{eq:22}) and (\ref{eq:23})) for $a$ and $b$ and taking the limit $c \rightarrow \infty$ result in the ratio $a / b = \sqrt{3}$, which is independent of the values of $v$ and $\mathrm{d}v / \mathrm{d}r$ at $r = r_0$. 
Therefore, the eigenfunction in the radiative zone can be given as $W_n = \sqrt{3} \mathrm{Ai} + \mathrm{Bi}$. 
Note that knowing the ratio $a/b$ is sufficient for evaluating the analytical expression of the $\Delta P_g$ pattern because the perturbed frequency $\Delta \omega_n$ is scale invariant in terms of $W_n$ (see Equation (\ref{eq:15})). 

Although we have specified the functional form of $W_n$ in quite an \textit{ad hoc} manner, the eigenfunctions thus computed can reproduce, e.g., the spatial phase of eigenfunctions that are computed by numerically solving Equation (\ref{eq:1}) for the toy BV frequencies constructed in the last section under the boundary conditions $v=0$ at the both center ($r=0$) and surface ($r=r_1$) (Figure \ref{fig:4}). 
The primary reason why the analytical expression of $W_n$ works well is that the eigenfunctions can be described by the Airy equation nearly throughout the radiative zone. 
In computing $W_n$ (the red curves in Figure \ref{fig:4}), the eigenfrequency $\omega_n$ is determined in a way that the location of the $n$-th node from the center in $\sqrt{3} \mathrm{Ai}(\zeta) + \mathrm{Bi}(\zeta)$ matches the surface ($r = r_1$ and $\zeta = \zeta_1$). 
We will use $W_n$ and $\omega_n$ thus computed in the next section as well. 

Finally, it should be instructive to mention that a common choice for $W_n$ is the JWKB solution that is obtained based on the asymptotic expansion of the Airy function: 
\begin{eqnarray}
	W_n \propto k_r^{-1/2} \sin \biggl ( \int_{r_0}^{r} \frac{LN}{\omega r'} \mathrm{d}r' + \frac{\pi}{4} + \delta \biggr ), \label{eq:24}
\end{eqnarray}
where $\delta$ represents the phase jump originating from the boundaries of the g-mode cavity (C19). 
It is often the case that the phase jump $\delta$ is treated as a free parameter which needs to be fine-tuned to reproduce the $\Delta P_g$ pattern (C19; H23). 
Interestingly, based on the asymptotic approximation of $\mathrm{Ai}$ and $\mathrm{Bi}$, it can be shown that 
\begin{eqnarray}
	\sqrt{3} \mathrm{Ai}(\zeta) + \mathrm{Bi}(\zeta) \sim \frac{2}{\sqrt{\pi}} |\zeta|^{-1/4} \sin \biggl ( \frac{2}{3} \zeta^{3/2} - \frac{\pi}{12} + \frac{\pi}{2}\biggr ),  \label{eq:25}
\end{eqnarray}
and the comparison between Equations (\ref{eq:24}) and (\ref{eq:25}) ends up with $\delta \sim \pi / 6$ (see also Equation (\ref{eq:4}) for the definition of $\zeta$). 
The phase jump may deviate from $\pi / 6$ in the presence of the convective overshooting that has been neglected in our discussions at the moment. 
In other words, however, such a deviation from $\pi / 6$ may imply that there should be an overshooting region around the convective boundary. 
Further discussions will be given in paper 2. 





\begin{figure}[t]
	\begin{center}
	\includegraphics[scale=0.47]{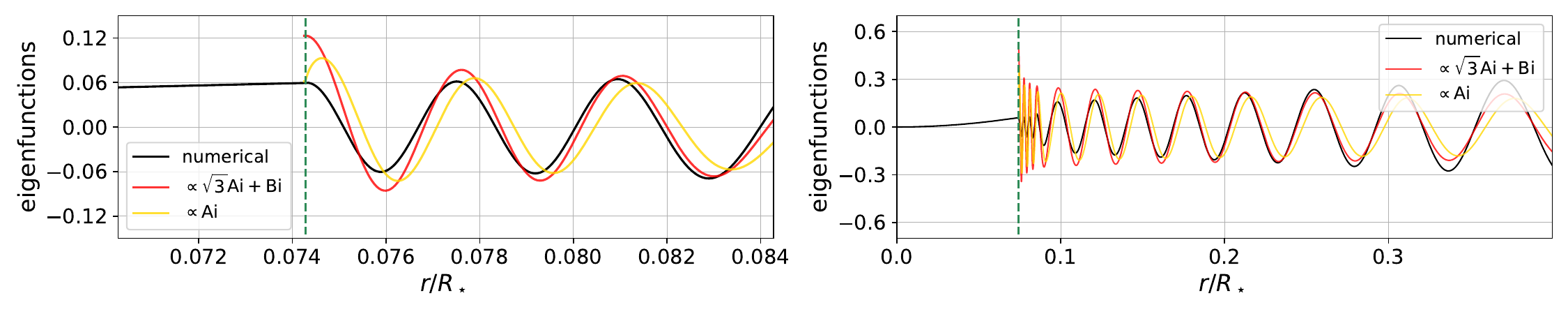}
		\caption{\footnotesize Comparison among the three kinds of eigenfunctions: the eigenfunction obtained by numerically solving Equation (\ref{eq:1}) where the toy BV frequency models are used (black), that proportional to $\sqrt{3} \mathrm{Ai} + \mathrm{Bi}$ (red), and that proportional to $\mathrm{Ai}$ (yellow), where $\mathrm{Ai}$ and $\mathrm{Bi}$ are the first and second kind of the Airy function. 
		The eigenfunctions are shown as a function of the stelar fractional radius $r/R_\star$, and the convective boundary is represented by the green dashed line. 
		The spherical degree and the radial order of the eigenfunctions are $n = -30$ and $\ell = 1$. 
		Because our first-order perturbative approach focuses only on the radiative zone, the analytical eigenfunctions (red and yellow) are not drawn in the convective core. 
		It is seen that the eigenfunction proportional to $\sqrt{3} \mathrm{Ai} + \mathrm{Bi}$ (red) can reproduce the spatial phase of the numerically computed eigenfunction (black) reasonably well not only around the convective boundary (the left panel) but also in a broader region of the radiative envelope (the right panel). 
		This is not the case between the numerical eigenfunction and that proportional to $\mathrm{Ai}$; the locations at which the eigenfunction is zero are different. 
		Note that the difference in the amplitude of the eigenfunctions are attributed to the fact that the numerical eigenfunction (black) is $v$ while the analytical eigenfunctions (red and yellow) are $W$ (remember the relation $v = (|\mathrm{d}r/\mathrm{d}\zeta|)^{1/2} W$). 
		Accordingly, we multiply the eigenfunctions with some suitable constants for visual aid. 
		The values of the constants are different for the left and right panels. 
		It should nevertheless be emphasized that the locations where the eigenfunction is zero do not change even if we consider the factor $(|\mathrm{d}r/\mathrm{d}\zeta|)^{1/2}$. 
		Thus, the phase match between the numerical eigenfunction and that proportional to $\sqrt{3} \mathrm{Ai} + \mathrm{Bi}$ is well founded. 
		}
	\label{fig:3}
	\end{center}
\end{figure}

\subsection{Results} \label{sec:3-3} 
In this section, using the toy models of the BV frequencies constructed in Section \ref{sec:3-1}, we carry out the validity check of the analytical expression for the $\Delta P_g$ pattern that is derived based on the first-order perturbation theory (Section \ref{sec:2}). 
To this end, we have computed two kinds of $\Delta P_g$ patterns. 
One is the ``numerical'' $\Delta P_g$ pattern; namely, we numerically solve the second-order differential equation (Equation (\ref{eq:1})) with the boundary conditions ($v= 0$ at the outer and inner edge of the mode cavity) to obtain the eigenperiods $P_n$, and we then take difference between two periods with the neighboring radial order $n$. 
We only consider the dipole modes ($\ell = 1$), and the range of the radial order is $-50 < n < -10$. 

The other one is the ``perturbative'' $\Delta P_g$ pattern that is computed based on the perturbation theory we developed in Section \ref{sec:2}. 
We start with computing $\zeta'$ by integrating $N$ of the toy model (Equation (\ref{eq:8})), with which we also computed the $f$-term. 
To isolate the $f_1$-term, we multiply the $f$-term by a window function that is one in a region where the first derivative of the BV frequency sharply transitions well inside the radiative zone; e.g., in the case of the toy model that mimics a stellar model with $(M, \, X_{\mathrm{c}}) = (1.6 \, M_\odot, \, 0.4)$ (the left panel of Figure \ref{fig:2}), the region is $0.08 < r/R_\star < 0.095$. 
The window function is zero in the other regions. 
The $f_0$-term is given by subtracting the $f_1$-term thus obtained from the $f$-term, i.e., $f_0 = f - f_1$. 
We have used $W_n$ and $\omega_n$ that are obtained in a manner explained in the last section. 
Equation (\ref{eq:15}) and $\Delta \omega_n = - 2 P_{n}^{-2} \Delta P_n$ together with Equation (\ref{eq:17}) allow us to compute the ``perturbative'' $\Delta P_g$ pattern. 

\begin{figure}[t]
	\begin{center}
	\includegraphics[scale=0.34]{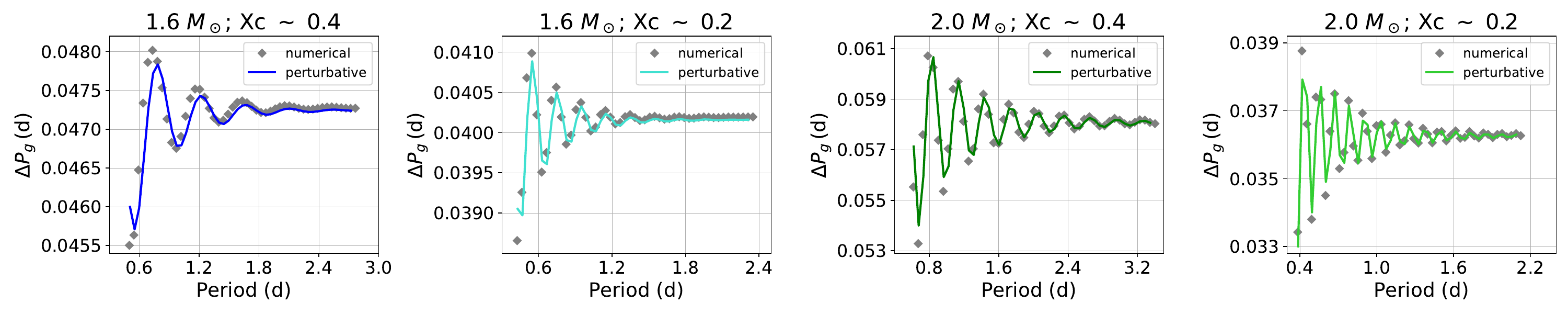}
		\caption{\footnotesize Comparison of the ``numerical'' (the grey diamonds) and the ``perturbative'' (the colored curves) $\Delta P_g$ patterns in the case of the toy BV frequencies that are constructed based on the four stellar models with different masses ($M = 1.6 \, M_\odot$ or $2 \, M_\odot$, blueish or greenish, respectively) and different evolutionary stages ($X_\mathrm{c} \sim 0.4$ or $0.2$, darker or lighter, respectively). 
		The g-mode period and $\Delta P_g$ are expressed in units of $\mathrm{d}$. 
		We see excellent agreements between the two types of $\Delta P_g$ patterns in terms of the both amplitude and phase of the oscillatory component in the $\Delta P_g$ patterns. }
	\label{fig:4}
	\end{center}
\end{figure}

Figure \ref{fig:4} compares the two kinds of $\Delta P_g$ patterns thus computed. 
We see reasonable agreements between the ``numerical'' (the grey diamonds) and the ``perturbative'' (the colored curves) $\Delta P_g$ patterns in all the cases, highlighting that the first-order perturbation theory (Section \ref{sec:2}), as well as the way we determine the explicit functional form of $W_n$ (Section \ref{sec:3-2}), can be fairly useful for describing the $\Delta P_g$ pattern in the case of the toy BV frequency models. 

It should especially be emphasized that the ``perturbative'' $\Delta P_g$ patterns successfully reproduce the period dependence of the amplitude in the ``numerical'' $\Delta P_g$ patterns. 
The decreasing amplitude of the $\Delta P_g$ pattern as a function of the g-mode period can be explained based on Equation (\ref{eq:15}). 
For a given $f_1$-term which is solely determined by the BV frequency profile and is independent of the mode frequency, the integration in Equation (\ref{eq:15}) is larger (smaller) for g modes with lower (higher) radial orders because $W_n$, that approximately is the trigonometric function, oscillates less (more) rapidly for lower-order (higher-order) g modes. 
In the limit of high radial orders, the integration in Equation (\ref{eq:15}) asymptotes to a constant that is frequency-independent, resulting in $\Delta P_g \sim \Delta P_{g,\mathrm{unptb}}$, which is identical to the result of the JWKB analysis of high-order g modes by, e.g., U89. 
The frequency dependence outside the integration in Equation (\ref{eq:15}) corresponds to that in the H23's semi-analytical expression, namely, $A_{\star} \propto P_g^{-1}$ (see H23 for more details); thus, the integration describes the finite-wavelength effect that was missing in the study of H23 (see also Section \ref{sec:intro}.) 
As we will see in paper 2, further analysis of Equation (\ref{eq:15}) leads us to a relationship between the g-mode period and the amplitude of the oscillatory component in the $\Delta P_g$ pattern, enabling us to revise H23's expression for the $\Delta P_g$ pattern. 


\section{Tests with the BV frequency profiles of stellar models} \label{sec:4}
In this section, we would like to focus on more realistic cases. 
Specifically, we use BV frequency profiles of stellar models instead of the toy models constructed in Section \ref{sec:3} and compare ``perturbative'' $\Delta P_g$ patterns with ``numerical'' ones in a way similar to those presented in Section \ref{sec:3}. 
A primary goal of the comparison is to obtain a rough picture about to what extent the first-order perturbation theory is applicable. 
Actually, as we will see, the perturbative approach is sometimes useful even for explaining $\Delta P_g$ patterns of intermediate-mass ($\sim 2 \, M_{\odot}$) main-sequence stellar models. 

We firstly explain how to calculate the ``numerical'' $\Delta P_g$ patterns. 
We used the same stellar models as those described in Section \ref{sec:3-1} whose linear adiabatic oscillation has been solved numerically with the community code GYRE \citep{2013MNRAS.435.3406T}. 
For the oscillation computations, the full fourth-order differential equation is solved with the default settings of GYRE. 
We focus on dipole ($\ell = 1$) high-order g modes with the radial order $-40 < n < -13$, which roughly corresponds to g modes typically observed in the case of $\gamma$ Dor and SPB stars \citep{2016MNRAS.455L..67M}. 
With the eigenperiods thus obtained, we directly computed the ``numerical'' $\Delta P_g$ patterns for each stellar model. 

The ``perturbative'' $\Delta P_g$ patterns are computed in the same way as we described in Section \ref{sec:3-2}; after we compute $\zeta'$ and the $f$-term based on the BV frequency of the corresponding stellar model, we use Equations (\ref{eq:15}) and (\ref{eq:17}) to evaluate the ``perturbative'' $\Delta P_g$ pattern. 
The $f_1$-term and the $f_0$-term are separated by imposing a suitable window function as demonstrated in Section \ref{sec:3-3}. 
How we separate the $f_0$- and $f_1$-terms does not strongly affect the oscillatory pattern in the ``perturbative'' $\Delta P_g$ pattern, but the general trend in the $\Delta P_g$ patterns can be affected (see discussions in the second last paragraph in this section.) 
For $\Delta P_{g,\mathrm{unptb}}$ in Equation (\ref{eq:17}), we have used the mean of the ``numerical'' $\Delta P_g$ pattern.  

\begin{figure}[t]
	\begin{center}
	\includegraphics[scale=0.34]{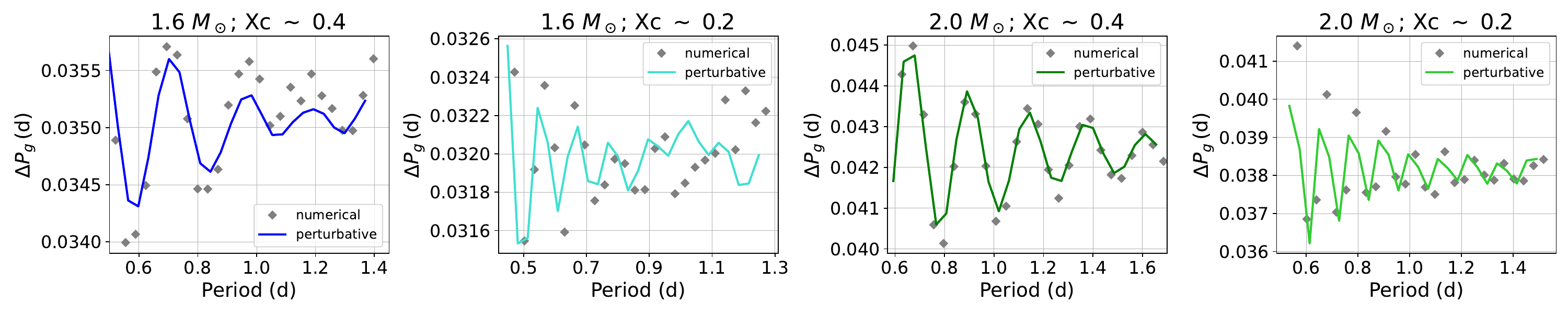}
		\caption{\footnotesize Same as Figure \ref{fig:4} except that the ``numerical'' (the grey diamonds) and ``perturbative'' (the colored curves) $\Delta P_g$ patterns are computed based on the BV frequency profiles of the stellar models. 
		We see more or less good agreements between the two kinds of $\Delta P_g$ patterns. }
	\label{fig:5}
	\end{center}
\end{figure}

Figure \ref{fig:5} shows the comparison between the ``numerical'' (the grey diamonds) and ``perturbative'' (the colored curves) $\Delta P_g$ patterns. 
Remembering that our perturbative approach is based on the second-order differential equation (Equation (\ref{eq:1})) meanwhile the ``numerical'' $\Delta P_g$ patterns are computed via GYRE that solves the forth-order differential equation, it might be surprising to see some agreements between the two kinds of $\Delta P_g$ patterns; the first-order perturbative approach can explain some of the oscillation computations conducted via GYRE. 
We especially see good agreements for the younger stellar models with $X_{\mathrm{c}} \sim 0.4$. 

\begin{figure}[t]
	\begin{center}
	\includegraphics[scale=0.47]{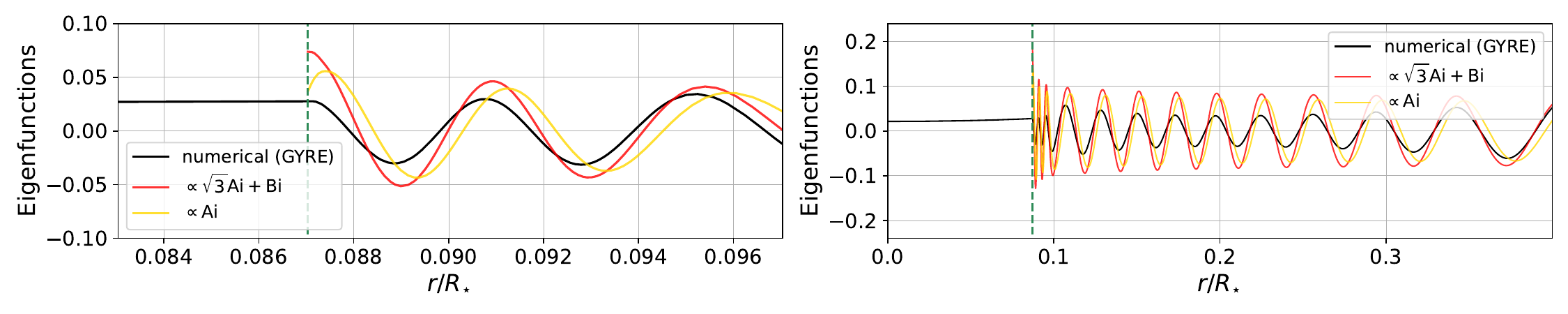}
		\caption{\footnotesize Same as Figure \ref{fig:3} except for the following two points: that the BV frequency profiles of the stellar models are used for the computation of the analytical eigenfunctions (red and yellow), and that the numerical eigenfunction (black) is computed by the oscillation code GYRE that numerically solves the fourth-order differential equation. 
		Even in the case of the stellar models, the eigenfunction proportional to $\sqrt{3} \mathrm{Ai} + \mathrm{Bi}$ can reasonably reproduce the spatial phase of the numerical eigenfunction. 
		Note that this might not be the case if we take into account the overshooting above the convective boundary (see discussions in Section \ref{sec:3-2}). 
		}
	\label{fig:6}
	\end{center}
\end{figure}

On one hand, the agreements are expected since 
we are currently focusing on the g modes with relatively high radial order ($n >$ few dozens); the Cowling approximation as well as the approximation by Equation (\ref{eq:2}), based on which our first-order perturbation theory is formulated, are appropriate for describing such high-order g modes. 
On the other hand, we would like to mention that 
the agreements are also attributed to the fact that the eigenfunctions numerically computed by GYRE are well approximated by those specified in Section \ref{sec:3-2}, i.e., $W_n \propto \sqrt{3} \mathrm{Ai} + \mathrm{Bi}$ (Figure \ref{fig:6}). 
Therefore, the g-mode sensitivity to the $f_1$-term is described (mostly) correctly. 
This would not be the case if the spatial phase of $W_n$ in Equation (\ref{eq:15}) was different from that of the corresponding eigenfunction computed via GYRE; the value of the integration in Equation (\ref{eq:15}) would vary, ending up with different $\Delta P_g$ patterns. 

Obviously, there is room for more work as well. 
We then would like to briefly discuss differences between the ``numerical'' and ``perturbative'' $\Delta P_g$ patterns, which is in particular prominent for the stellar model with $(M,\, X_\mathrm{c}) = (1.6 \, M_{\odot}, \, 0.2)$ (see the second left panel in Figure \ref{fig:5}.) 
The deviations may originate from numerical problems rather than a failure of the Cowling approximation because we see relatively larger deviations in the longer g-mode periods where the Cowling approximation should be appropriate. 
A possible main numerical issue is that the $f$-term profile is fairly dependent on numerical schemes of differentiation as is evident in the shaky behaviors of the $f$-terms (see the grey dashed curves in the bottom panels of Figure \ref{fig:2}). 
Such shaky behaviors in the $f$-terms are due to the fact that the $f$-term corresponds to the third derivative of, e.g., the mean molecular weight \citep[see Equation (\ref{eq:5}) and Equation (2) in][]{2008MNRAS.386.1487M}, and thus, it is numerically challenging for us to obtain a smooth $f$-term even though we somehow smoothen the chemical composition profile in a stellar model as we did in this study (Section \ref{sec:2-2}). 
In addition, in the case of stellar BV frequencies, the $f_0$- and $f_1$ terms are not so easily separable that the integration in Equation (\ref{eq:15}) should be dependent on what width we choose for the window function to decompose the $f$-term. 
Actually, when we take a look at the case of $1.6 \, M_{\odot}$ model with $X_\mathrm{c} = 0.2$, the ``numerical'' $\Delta P_g$ pattern exhibits a quasi-linear trend (see around $P_g > 0.9 \, \mathrm{d}$ in the second left panel of Figure \ref{fig:5}). 
As discussed in Section \ref{sec:2-3}, this quasi-linear trend probably is an oscillatory component with a long period ($n > 100$) that is caused by sharp structural variations just above the convective boundary. 
We can take into account the effect of sharp structural variations around the convective boundary on the ``perturbative'' $\Delta P_g$ pattern by, e.g., using broader window functions to separate $f_0$- and $f_1$-terms, but again, we would suffer from shaky behaviors of the $f$-term which hampers us from accurately evaluating the integration in Equation (\ref{eq:15}). 

Since our primary goal in this section is to roughly check if the perturbative approach could explain the ``numerical'' $\Delta P_g$ patterns in the case of stellar models, we here do not continue investigating the numerical issues that cause the deviations between the ``perturbative'' and ``numerical'' $\Delta P_g$ patterns. 
Nevertheless, it should be emphasized that we would benefit substantially from more thorough comparison of the numerical schemes adopted in our perturbative approach (central differentiation) and those used in GYRE. 
This is primarily because, once we confirm that the first-order perturbative approach is indeed useful for describing the $\Delta P_g$ pattern of stellar models, we can move on to inverting observed $\Delta P_g$ patterns to infer the BV frequency profile, or more correctly, the $f$-term \citep[structure inversion; e.g.,][]{1991sia..book..519G}. 
At the moment, carrying out structure inversion for the intermediate-mass main-sequence g-mode pulsators is considered to be difficult since we often suffer from strong non-linearity \citep{2023A&A...675A..17V}; in short, reference stellar models are so far from representing the corresponding stars that the frequency differences between the model and star cannot be described by the first-order perturbation theory. 
The existence of the strong non-linearity in structure inversion has motivated some groups to develop non-linear inversion methods \citep[e.g.,][]{2024A&A...686A.267F}. 
In addition to the non-linearity, we also have to consider effects of fast rotation that characterizes the intermediate-mass main-sequence stars. 
In this context, our framework might help us to carry out structure inversion in a \textit{linear} regime because we do not need a reference stellar model in our formulation, and all we have to do is to determine the mode indices, $(n, \, \ell, \, m)$ (mode identification). 
In addition, the rotational effects can be taken into account via TAR \citep{eckart1960hydrodynamics,1997ApJ...491..839L}. 
A promising approach for inverting observed $\Delta P_g$ patterns based on the perturbative formulation has recently discussed by \citet{2025arXiv251105780G}. 
We will investigate the possibility of linear structure inversion based on our formulation in the forthcoming paper (not in this series), which can complement other studies of the $\Delta P_g$ patterns, such as the development of non-linear inversion methods \citep[e.g.,][]{2024A&A...686A.267F}, the direct stellar modeling in 1-d/2-d setups \citep[e.g.,][]{2024A&A...685A..21M,2023A&A...677L...5M}, and the derivation of the semi-analytical expression of the $\Delta P_g$ pattern \citep[][C19, and paper 2]{2015ApJ...805..127C}.

\section{Conclusion and prospects} \label{sec:5}
Perturbative approaches are advantageous in that we can evaluate the effects of any arbitrary shape of the BV frequency transition on the $\Delta P_g$ pattern. 
However, if we favor the first-order perturbative analysis, perturbations should be small enough for us to neglect the terms higher than the second-order, which is usually not the case for ordinary stellar models that have a contrast ($\sim$ a few) between the peak and ground values of the BV frequency. 
Inspired by U89's formulation, we treat the $f_1$-term, which corresponds to a bump in the second derivative of the BV frequency, as a small perturbation 
to obtain the analytical expression of the $\Delta P_g$ pattern. 
According to the analytical expression, the amplitude of the oscillatory component in the $\Delta P_g$ pattern can be interpreted as an average of the $f_1$-term weighted by the sensitivity kernel; this weighted averaging actually represents the finite-wavelength effect which H23 has neglected. 
We have specified that the sensitivity kernel in the radiative envelope is proportional to $\sqrt{3} \mathrm{Ai} + \mathrm{Bi}$ when we assume no overshooting beyond the convective boundary. 
The analytical expression can reproduce the $\Delta P_g$ patterns very well in the case of the toy BV frequency models. 
In the case of stellar models (with the mass of $\sim 2 \, M_\odot$), the agreements are less prominent compared with the case of the toy models, but the phase of the $\Delta P_g$ patterns is nicely reproduced. 
Overall, the first-order perturbative approach presented in this study may be useful for describing the $\Delta P_g$ pattern of main-sequence g-mode pulsators with the mass of $\sim 2 \, M_\odot$. 

The success of the perturbation theory in explaining the $\Delta P_g$ pattern of stellar models could be a good news for inversion players. 
Since our formulation does not need a reference stellar model (although we have to conduct mode identification in advance), we may not suffer from strong non-linearity in carrying out structure inversion. 
The effects of fast rotation, which is typical for the early-type main-sequence stars, can be taken into account via the traditional approximation of rotation (TAR). 
We may therefore carry out structure inversion of the $\Delta P_g$ pattern in a linear regime based on the first-order perturbative formulation we have established in this study. 
Nevertheless, the rough comparison between the ``perturbative'' and ``numerical'' $\Delta P_g$ patterns we have shown in this study is not sufficient for us to conclude that the first-order perturbative approach can be useful for structure inversion.  
The next step would be then a more thorough comparison between the ``perturbative'' and ``numerical'' $\Delta P_g$ patterns. 

Another promising direction may be the semi-analytical approach. 
%
As will be discussed in paper 2, detailed comparison between the analytical expression of the $\Delta P_g$ pattern derived in this study and the semi-analytical expression derived by H23 enables us to explicitly write down the period dependence of the amplitude of the oscillatory component in the $\Delta P_g$ pattern. 
Using the explicit period dependence, we can derive a new semi-analytical expression for the $\Delta P_g$ pattern that can reproduce the $\Delta P_g$ patterns of lower-mass ($\sim 2 \, M_\odot$) stellar models (paper 2). 
Furthermore, 
we would easily implement the traditional approximation of rotation (TAR) in the new semi-analytical expression; thus, we will possibly be able to explain the $\Delta P_g$ pattern of fast-rotators with relatively low masses, that correspond to $\gamma$ Dor stars. 
These semi-analytical studies would also be complementary to other studies of the $\Delta P_g$ patterns, such as non-linear structure inversion and the 1-d/2-d stellar modeling, thus providing us with a precious opportunity to infer the deep interiors of g-mode pulsators from different perspectives. 
%

\section*{Acknowledgements}
Y.H. and T.S. acknowledge M. Takata, O. Benomar, and T. Tokuno for their constructive comments. 
D. Reese, R.-M. Ouazzani, G. Buldgen, and M.-A. Dupret are also thanked for their insightful discussions. 
Y.H. and T.S. would also like to thank the anonymous referee for the insightful comments.
This work was supported by JSPS Grants-in-Aid for JSPS fellows Grant No. JP23KJ0300, JSPS Grants-in-Aid for Early-Career Scientists Grant No. JP24K17087, and JSPS Grants-in-Aid for Scientific Research (B) Grant No. JP24K00654. 

\bibliography{DPg_HS24_1_rev1}{}
\bibliographystyle{aasjournal}

\end{document}